\documentclass[graybox]{svmult}

\usepackage{makeidx}         
\usepackage{graphicx}        
\usepackage{multicol}        
\usepackage[bottom]{footmisc}

\usepackage{newtxtext}       %
\usepackage{newtxmath}       

\makeindex             

\begin{document}

\title*{Energy inequalities in interacting quantum field theories}
\author{Daniela Cadamuro}
\institute{Daniela Cadamuro \at Universit\"at Leipzig, Institut f\"ur Theoretische Physik, Br\"uderstra\ss e 16, 04103 Leipzig, \email{daniela.cadamuro@itp.uni-leipzig.de}}
%
%
\maketitle

\abstract{
The classical energy conditions, originally motivated by the Penrose-Hawking singularity theorems of general relativity, are violated by quantum fields. A reminiscent notion of such conditions are the so called quantum energy inequalities (QEIs), which are however not known to hold generally in quantum field theory. Here we present first steps towards investigating QEIs in quantum field theories with self-interaction.}

\section{Introduction}
\label{sec:1}


One of the fundamental observables both in quantum and in classical field theory is the stress-energy
tensor $T^{\alpha \beta}$. It has a special role in general relativity, as the Einstein equations link the curvature of
spacetime to the distribution of matter throughout it. Certain positivity conditions on the stress-energy
tensor (e.g., the so called ``weak energy condition''), which are fulfilled by many classical matter fields,
imply severe constraints on exotic spacetime geometries. They also enter the Penrose and Hawking
singularity theorems \cite{HawPen1970}, positive mass theorems \cite{SchoenYau:1979}, and Hawking's chronology protection results \cite{Hawking:1992},
among many others.
However, in quantum field theory (QFT) these energy conditions are violated; the energy density can
have negative expectation values. This raises the question whether the energy conditions on the matter
in the assumptions of the singularity theorems are compatible with quantum matter, or whether quantum
fields allow the existence of ``exotic'' spacetimes like time machines, wormholes, warp drives.
To exclude these scenarios, there must be constraints to the extent in which quantum fields can cause
negative energy densities. These constraints are called ``Quantum Energy Inequalities'' (QEIs). First investigated
by Ford \cite{Ford:1991}, they are reminiscent of classical energy conditions, suggesting that the singularity
theorems can still hold for realistic matter \cite{FewsterGalloway:2011}.

These inequalities formally go over into the Averaged Null Energy Condition (ANEC) \cite{Flanagan:1996}
when the stress-energy tensor is averaged over a null geodesic. The ANEC has
recently received considerable attention in the context of holography due to its relation with the quantum
information carried by black hole horizons (see, e.g., \cite{Fu:2017,Bousso:2015,Bousso16}).


Violations of the classical energy conditions must exist in any quantum field theory, since the vacuum expectation value of the energy density is supposed to vanish. For free quantum fields, it is straightforward to construct examples of states where the energy density is locally negative. For free scalar bosons this property does not arise by evaluating the energy density in fixed $n$ particle states, but by considering superpositions, for example, of the vacuum and a two-particle state. Namely, local negativity of the energy density appears as a quantum interference effect, and does not occur in the classical regime.  As another example of its intimate relation to quantum effects, consider the Casimir effect where an attractive force is generated between infinitely extended parallel plates in the vacuum. One can explicitly compute the stress-energy tensor of the electromagnetic field and find that the energy density between the plates is negative, and depends on the inverse distance between the plates.

In a quantum field theory with interacting particles the situation is more involved. One finds that in certain models with interacting bosons, the energy density can be negative also in states of fixed particle number. Namely, in the class of lower dimensional \emph{quantum integrable models}, one can find one-particle expectation values with locally negative energy density \cite{BostelmannCadamuroFewster:2013,BostelmannCadamuro:2016}. Hence, negativity of the energy density is more \emph{profound} in theories with interaction, and existence of QEIs is far from obvious and poses a challenging question.

Integrable models are a special class of $1+1$-dimensional QFTs, where the two-particle scattering process characterizes the theory completely. They are constructed as an inverse scattering problem, specifically, given a function $S_2$ as a mathematical input, one constructs the corresponding QFT having this two-particle scattering function.

There are several simplifications when considering integrable theories. First of all, unlike most interacting field theories they can be represented on a Fock space, with the interacting vacuum given by the usual Fock vacuum, allowing for explicit computations. Further, they are amenable to a treatment in a non-perturbative setting \cite{Lechner:2008}, avoiding to deal with formal power series whose convergence is generally unknown. 

Physically, they are \emph{toy} models for interaction, but share interesting common features with interacting theories in higher dimensions. For example, the nonlinear $O(N)$ sigma models \cite{BabujianFoersterKarowski:2013} are linked to experimentally realizable situations in condensed matter systems. They can also be regarded as simplified analogues of four-dimensional nonabelian gauge theories, inasmuch as they share crucial features with them, including renormalizability, asymptotic freedom, and the existence of instanton solutions.

Then there are models that can support a representation of a gauge group or a group of internal symmetries, for example, the 1+1 dimensional $SU(N)$-symmetric models (such as the chiral Gross-Neveu and principal chiral models) which are of importance in physics as toy models of quantum field theory with relations to both gauge and string theory, and have been analyzed both in general and in the limit $N \to \infty$. 

Finally, the Ising model is widely known also for its counterpart in statistical mechanics in the context of lattice spin systems. 

Therefore, integrable systems provide a \emph{``landscape''} of possible interactions, where one may hope to obtain hints for the abstract conditions that underlie the phenomenon of QEIs. 

Mathematically, QEIs are lower bounds for the smeared energy density, $T^{00}(g^2) =\int dt\; g^2(t)T^{00}(0,t) $ of the form
\begin{equation}\label{qei}
\langle  \varphi, T^{00}(g^2) \varphi  \rangle \geq -c_g \| \varphi \|^2
\end{equation}
for all suitably regular state vectors $\varphi$ and all real-valued test functions $g$, where $c_g>0$ is a constant depending only on  $g$.  However this inequality may not hold in all physical applications, e.g., in the non-minimally coupled scalar field in a curved spacetime only a weaker form of this inequality can hold \cite{FewsterOsterbrink2008}, where the constant $c_g$ in Eq.~\eqref{qei} may depend on the total energy of the state.

QEIs of the form \eqref{qei} have been proved for the linear scalar field, linear Dirac field, linear vector field, both on flat and curved spacetime, the Rarita-Schwinger field, and for $1+1$ dimensional conformal fields (see \cite{ Few:qft_inequalities} for a review). 
Weaker forms of quantum inequalities have been proved for certain ``classically positive'' expressions in \cite{BostelmannFewster:2009}, but without a clear relation to the energy density. 

Only recently, a state-independent QEI has been established for the massive Ising model \cite{BostelmannCadamuroFewster:2013}, which represents the first result to our knowledge of a QEI in a self-interacting situation. Partial results have been obtained later in a larger class of ``scalar'' integrable systems, including the $\sinh$-Gordon model \cite{BostelmannCadamuro:2016}. 

In the next sections we will summarize the results of \cite{BostelmannCadamuroFewster:2013} and \cite{BostelmannCadamuro:2016}.

\section{QEIs in integrable systems at one-particle level}\label{sec:2}

In this section we summarize some results on QEIs in integrable systems with one species of scalar bosons. Some of these, for example, the $\sinh$-Gordon model \cite{FringMussardoSimonetti:1993}, can be derived from a classical Lagrangian, and a candidate for the energy density can be computed directly from the Lagrangian. Other models in this class, for example the generalized $\sinh$-Gordon model in Table 1 of \cite{BostelmannCadamuro:2016}, are not associated with a Lagrangian, and it is therefore not a priori clear what one should regard as the stress-energy tensor.

Our first task is therefore to find an intrinsic characterization of the stress-energy tensor $T^{\alpha \beta}$ from the generic properties of this observable, and independent of the specific form of scattering matrix.

We focus our attention on one-particle matrix elements of $T^{\alpha \beta}$, which is generically given as an integral kernel operator:
\begin{equation}
\langle \varphi, T^{\alpha \beta}(g^2) \psi \rangle = \int d\theta d\eta\; \overline{\varphi(\theta)}F^{\alpha \beta}(\theta,\eta)\psi(\eta),
\end{equation}
where $\varphi, \psi$ are vectors in the single-particle space $L^2(\mathbb{R})$.

The generic integral kernel $F^{\alpha \beta}$ is restricted by various properties of the stress-energy tensor: locality of the field 
$T^{\alpha \beta}(t,x)$, symmetry of the tensor $T^{\alpha \beta}$, covariance under Poincar\'e transformations and spacetime reflections, the continuity equation ($\partial_{\alpha}T^{\alpha \beta}=0$), and the fact that the $(0,0)$-component of the tensor integrates to the Hamiltonian ($\int dx\; T^{00}(t,x) = H$). We can show that these requirements are necessary and sufficient for $F^{\alpha \beta}$ to have the form \cite[Prop.~3.1]{BostelmannCadamuro:2016}:
\begin{equation}\label{coeff}
F^{\alpha \beta}(\theta,\eta) = F^{\alpha \beta}_{\text{free}}(\theta, \eta)
\underbrace{P(\cosh(\theta -\eta))F_{\text{min}}(\theta -\eta +i\pi)}_{=:F_P(\theta-\eta)}
\widetilde{g^2} (\mu\cosh\theta-\mu\cosh\eta),
\end{equation}
where the individual factors are as follows: $P$ is a real polynomial with $P(1)=1$, $\mu>0$ is the mass of the particle and $\sim$ denotes Fourier transform. $
F^{\alpha \beta}_{\text{free}}$ stands for the expression of the ``canonical'' stress-energy tensor of the free Bose field,
\begin{equation}
F^{\alpha \beta}_{\text{free}}(\theta,\eta) = \frac{\mu^2}{2\pi}
\left( \begin{array}{ccc}
\cosh^2\big( \frac{\theta +\eta}{2}\big) & \frac{1}{2}\sinh (\theta +\eta ) \\
\frac{1}{2}\sinh(\theta +\eta) & \sinh^2 \big( \frac{\theta +\eta}{2} \big)  \end{array} \right).
\end{equation}
$F_{\text{min}}$ is the so called \emph{minimal solution} of the model \cite{KarowskiWeisz:1978}, which encodes the dependence on the scattering function $S_2$ in a unique way. For the free field, $F_{\text{min}}(\zeta)=1$ and for the Ising model, $F_{\text{min}}(\zeta)=-i\sinh \frac{\zeta}{2}$; for the sinh-Gordon model, see \cite{FringMussardoSimonetti:1993}.

The function $F_P$ in Eq.~\eqref{coeff} now determines the negativity of the energy density and \mbox{(non-)}existence of QEIs.

\emph{\bf Negative values of the energy density.}
If there is a $\theta_P \in \mathbb{R}$ such that $|F_P(\theta_P)|>1$, then there exists a one-particle state $\varphi \in L^2(\mathbb{R})$ and a real-valued Schwartz function $g$ such that $\langle \varphi, T^{00}(g^2) \varphi \rangle < 0$ \cite[Prop.~4.1]{BostelmannCadamuro:2016}. 
 
As one can see from the above examples of $F_{\text{min}}$, this is fulfilled in the case of the Ising and $\sinh$-Gordon models, but not for the free field if $P=1$. In fact, it is known that for free bosons the one-particle energy density is positive.

\emph{\bf Existence of QEIs.}
The existence of QEIs is determined by the behaviour of $F_P$ for large arguments. Namely, if $|F_P(\zeta)| \leq c \cosh \operatorname{Re} \zeta$ in a small strip around the real axis and with a constant $0<c< \frac{1}{2}$, then 
\begin{equation}\label{oneqei}
\forall \varphi \in \mathcal{D}(\mathbb{R}): \quad \langle \varphi, T^{00}(g^2) \varphi \rangle \geq - c_g \| \varphi \|^2
\end{equation}
for all real-valued Schwartz function $g$. That is, a state-independent QEI exists at least at one-particle level \cite[Thm.~5.1]{BostelmannCadamuro:2016}.

If, on the other hand, $F_P(\theta) \geq c \cosh\theta$ for $c > \frac{1}{2}$ and for large $\theta$, then an inequality of the type \eqref{oneqei} cannot hold \cite[Proposition~4.2]{BostelmannCadamuro:2016}. This is a \emph{no-go} theorem on existence of QEIs.

\emph{\bf Form of the energy density.}
We can see from the above that requiring a QEI to exists restricts the form of the energy density, namely the choice of $P$, sometimes fixing it uniquely.

In particular, in the Ising model, where $F_{\text{min}}(\theta +i\pi)$ grows like $\cosh \frac{\theta}{2}$ at large values of $\theta$, a QEI holds if and only if $P \equiv 1$.

Instead, in the free and sinh-Gordon models, $F_{\text{min}}$ converges to a constant for large $\theta$, thus a QEI can hold only if $\text{deg }P=0,1$. This means that we are left with the choice
\begin{equation}
P(x)=(1-\alpha) +\alpha x \quad \text{with } \alpha \in \mathbb{R},\;  |\alpha|< \frac{1}{2F_{\text{min}}(\infty +i\pi)},
\end{equation}
where $F_{\text{min}}(\infty +i\pi) := \lim_{\theta \rightarrow \infty}F_{\text{min}}(\theta +i\pi)$, and therefore the choice is at least strongly restricted.

\section{Special example: The Ising model}

In the massive Ising model one can prove a stronger result, namely, that a state-independent QEI exists which is independent of the particle number \cite{BostelmannCadamuroFewster:2013}. As already remarked, this represents the first proof to our knowledge of a QEI in the form \eqref{qei} in a self-interacting QFT.

The massive Ising model is the simplest interacting example of an integrable quantum field theory, where the two-particle $S$-matrix is given by $S_2 =-1$. Although it is related to a free Majorana field, it is an interacting theory of scalar bosons, namely, the asymptotic incoming and outgoing scattering states are bosonic. 

The energy density of this theory also behaves subtly different from the free bosonic case: As mentioned in Section~\ref{sec:2}, there are single-particle states of the Ising model where the energy density is locally negative. Moreover, the energy density does not scale with the particle number; when considering incoming $n$-particle states of the form $\varphi_n = \varphi_1 \otimes_{\text{in}} \cdots \otimes_{\text{in}} \varphi_1$ we have in general
\begin{equation}
\langle \varphi_n, T^{00}(g) \varphi_n \rangle \neq n \langle \varphi_1, T^{00}(g) \varphi_1 \rangle
\end{equation}
whereas in a free bosonic theory, equality would hold.

The theory is formulated on a fermionic Fock space in terms of a wedge local field $\phi(\pmb{x})$. On the same Fock space one can also consider a free Majorana field $\psi(\pmb{x})$. These two fields are not relatively local, and while they share the same energy-momentum operators, they have different PCT operators. Bilocal expressions in $\psi$ are local observables in the Ising model, but there are many other local observables in the Ising model (including local fields~\cite{SchroerTruong:1978}) that do not arise in this way. Despite these differences, it turns out that the energy densities of the two fields coincide, namely
\begin{equation}
T^{00}(\pmb{x}) = \frac{i}{4} {:} \psi(\pmb{x})^T \partial_t\psi(\pmb{x}) - (\partial_t\psi(\pmb{x}))^T \psi(\pmb{x}) {:} \, 
\end{equation}
which is relatively local to both. (This corresponds to $P=1$ in \eqref{coeff}.)

Now we can use methods developed for the Dirac fields in $3+1$ dimensional spacetime \cite{FewsterMistry:2003,Dawson:2006} to establish a state-independent quantum energy inequality of the form \eqref{qei} valid on the entire Hilbert space.
Specifically, we obtain
\begin{equation}\label{eq:QEI}
\smash{\int} dt\,  g(t)^2 \langle \varphi, T^{00}(t,x) \varphi \rangle  \geq -\frac{1}{4\pi^2}\smash{\int_\mu^\infty} \!\!\!\! 
d\omega\, \omega^2
|\widetilde{g}(\omega)|^2 Q \Big(\frac{\omega}{\mu}\Big) \| \varphi \|^2,
\end{equation}
for any real-valued smooth compactly supported function $g$
and all sufficiently regular states $\varphi$, where
the function $Q:[1,\infty)\to \mathbb{R}^+$
is given by
\begin{equation}\label{eq:Qpm}
Q(u) = \sqrt{1-u^{-2}}- u^{-2}\log(u+\sqrt{u^2-1}).
\end{equation}
As a comparison with the massless case shows \cite{BostelmannLechnerMorsella:2011,FewsterHollands:2005}, where the theory becomes conformal, the constant on the right hand side of \eqref{eq:QEI} is not optimal.

\section{Conclusions}

We have shown that a state-independent QEI exists for the simplest example of an integrable QFT, namely the massive Ising model. For a larger class of integrable models, QEIs exists at least at one-particle level. The situation for general Fock states is unclear at the moment, but we expect QEIs to hold for all particle numbers as suggested by numerical evidence at two-particle level.

Even if our examples are quite simplistic, they already indicate some fundamental features of the energy density and of QEIs. 
QEIs are often considered to be related to stability of spacetime, and as a localized version of the positivity of energy condition, i.e., thermodynamical stability. However, our results indicate that QEIs may be important as a selection criteria for the energy density itself. Already at one-particle level they exclude a number of choices that are compatible with the usual covariance and locality requirements, in some cases even fixing it uniquely.

It is also interesting to note that even in the simplest model of interaction the energy density can distinguish between a free and an interacting theory; for example, the massive Ising model allows for negative energy density in one-particle states in contrast to a free bosonic theory, even if the two theories look identical at asymptotic times.

Clearly, one would be interested to investigate these features in other interacting field theories. The larger class of integrable models may provide suitable examples, such as models with several particle species, with gauge symmetries and with bound states. Other accessible models are the $P(\phi)_2$-models on Minkowski and de Sitter space, which have been constructed non-perturbatively; these would be particularly interesting because of their feature of particle production, which is absent in integrable theories.

\section*{Acknowledgements}

The author is supported by the Deutsche Forschungsgemeinschaft (DFG) within the Emmy Noether grant CA1850/1-1. 


\bibliographystyle{abbrv} 
\bibliography{integrable}

\end{document}